# Optimization of Trading Physics Models of Markets

Lester Ingber

Lester Ingber Research

POB 06440 Sears Tower, Chicago, IL 60606

and

DRW Investments LLC

311 S Wacker Dr, Ste 900, Chicago, IL 60606

ingber@ingber.com, ingber@alumni.caltech.edu

and

Radu Paul Mondescu

DRW Investments LLC

311 S Wacker Dr, Ste 900, Chicago, IL 60606

rmondescu@drwtrading.com

## ABSTRACT

We describe an end-to-end real-time S&P futures trading system. Inner-shell stochastic nonlinear dynamic models are developed, and Canonical Momenta Indicators (CMI) are derived from a fitted Lagrangian used by outer-shell trading models dependent on these indicators. Recursive and adaptive optimization using Adaptive Simulated Annealing (ASA) is used for fitting parameters shared across these shells of dynamic and trading models.

# 1. INTRODUCTION

## 1.1. Approaches

Real-world problems are almost intractable analytically, yet methods must be devised to deal with this complexity to extract practical information in finite time.  This is indeed true in the field of financial engineering, where time series of various financial instruments reflect nonequilibrium, highly non-linear, possibly even chaotic [1] underlying processes.  A further difficulty is the huge amount of data necessary to be processed.  Under these circumstances, to develop models and schemes for automated, profitable trading is a non-trivial task.

In the context of this paper, it is important to stress that dealing with such complex systems invariably requires modeling of dynamics, modeling of actions on these dynamics, and algorithms to fit parameters in these models to real data.  We have elected to use methods of mathematical physics for our models of the dynamics, artificial intelligence (AI) heuristics for our models of trading rules acting on indicators derived from our dynamics, and methods of sampling global optimization for fitting our parameters.  Too often there is confusion about how these three elements are being used for a complete system.  For example, in the literature there often is discussion of neural net trading systems or genetic algorithm trading systems.  However, neural net models (used for either or both models discussed here) also require some method of fitting their parameters, and genetic algorithms must have some kind of cost function or process specified to sample a parameter space, etc.

Some powerful methods have emerged during years, appearing from at least two directions: One direction is based on inferring rules from past and current behavior of market data leading to learning-based, inductive techniques, such as neural networks, or fuzzy logic.  Another direction starts from the bottom-up, trying to build physical and mathematical models based on different economic prototypes.  In many ways, these two directions are complementary and a proper understanding of their main strengths and weaknesses should lead to synergetic effects beneficial to their common goals.

Among approaches in the first direction, neural networks already have won a prominent role in the financial community, due to their ability to handle large quantities of data, and to uncover and model nonlinear functional relationships between various combinations of fundamental indicators and price data [2,3].

In the second direction we can include models based on non-equilibrium statistical mechanics [4] fractal geometry [5], turbulence [6], spin glasses and random matrix theory [7], renormalization group [8], and gauge theory [9]. Although the very complex nonlinear multivariate character of financial markets is recognized [10], these approaches seem to have had a lesser impact on current quantitative finance practice, although it is becoming increasing clear that this direction can lead to practical trading strategies and models.

To bridge the gap between theory and practice, as well as to afford a comparison with neural networks techniques, here we focus on presenting an effective trading system of S&P futures, anchored in the physical principles of nonequilibrium statistical mechanics applied to financial markets [4,11].

Starting with nonlinear, multivariate, nonlinear stochastic differential equation descriptions of the price evolution of cash and futures indices, we build an algebraic cost function in terms of a Lagrangian. Then, a maximum likelihood fit to the data is performed using a global optimization algorithm, Adaptive Simulated Annealing (ASA) [12]. As firmly rooted in field theoretical concepts, we derive market canonical momenta indicators, and we use these as technical signals in a recursive ASA optimization that tunes the outer-shell of trading rules. We do not employ metaphors for these physical indicators, but rather derive them directly from models fit to data.

The outline of the paper is as follows: Just below we briefly discuss the optimization method and momenta indicators. In the next three sections we establish the theoretical framework supporting our model, the statistical mechanics approach, and the optimization method. In Section 5 we detail the trading system, and in Section 6 we describe our results. Our conclusions are presented in Section 7.

## 1.2. Optimization

Large-scale, non-linear fits of stochastic nonlinear forms to financial data require methods robust enough across data sets. (Just one day, tick data for regular trading hours could reach 10,000–30,000 data points.) Simple regression techniques exhibit deficiencies with respect to obtaining reasonable fits. They too often get trapped in local minima typically found in nonlinear stochastic models of such data. ASA is a global optimization algorithm that has the advantage — with respect to other global optimization methods as genetic algorithms, combinatorial optimization, etc. — not only to be efficient in its importance-sampling search strategy, but to have the statistical guarantee of finding the best optima [13,14]. This gives some confidence that a global minimum can be found, of course provided care is taken as necessary to tune the algorithm [15].

It should be noted that such powerful sampling algorithms also are often required by other models of complex systems than those we use here [16]. For example, neural network models have taken advantage of ASA [17-19], as have other financial and economic studies [20,21].

### 1.3. Indicators

In general, neural network approaches attempt classification and identification of patterns, or try forecasting patterns and future evolution of financial time series. Statistical mechanical methods attempt to find dynamic indicators derived from physical models based on general principles of non-equilibrium stochastic processes that reflect certain market factors. These indicators are used subsequently to generate trading signals or to try forecasting upcoming data.

In this paper, the main indicators are called Canonical Momenta Indicators (CMI), as they faithfully mathematically carry the significance of market momentum, where the "mass" is inversely proportional to the price volatility (the "masses" are just the elements of the metric tensor in this Lagrangian formalism) and the "velocity" is the rate of price changes.

## 2. MODELS

### 2.1. Langevin Equations for Random Walks

The use of Brownian motion as a model for financial systems is generally attributed to Bachelier [22], though he incorrectly intuited that the noise scaled linearly instead of as the square root relative to the random log-price variable. Einstein is generally credited with using the correct mathematical description in a larger physical context of statistical systems. However, several studies imply that changing prices of many markets do not follow a random walk, that they may have long-term dependences in price correlations, and that they may not be efficient in quickly arbitraging new information [23-25]. A random walk for returns, rate of change of prices over prices, is described by a Langevin equation with simple additive noise $\eta$, typically representing the continual random influx of information into the market.

$$\dot{M} = -f + g\eta \, ,$$

$$\dot{M} = dM/dt \, ,$$

$$< \eta(t) >_\eta = 0 \, , \; < \eta(t), \eta(t') >_\eta = \delta(t - t') \, , \tag{1}$$

where $f$ and $g$ are constants, and $M$ is the logarithm of (scaled) price, $M(t) = \log(P(t)/P(t - dt))$. Price, although the most dramatic observable, may not be the only appropriate dependent variable or order parameter for the system of markets [26]. This possibility has also been called the "semistrong form of the efficient market hypothesis" [23].

The generalization of this approach to include multivariate nonlinear nonequilibrium markets led to a model of statistical mechanics of financial markets (SMFM) [11].

## 2.2. Adaptive Optimization of $F^x$ Models

Our S&P model for the futures $F$ is

$$dF = \mu \, dt + \sigma F^x dz \, ,$$

$$< dz > = 0 \, ,$$

$$< dz(t) \, dz(t') > = dt \, \delta(t - t')$$

We have used this model in several ways to fit the distribution's volatility defined in terms of a scale and an exponent of the independent variable [4].

A major component of our trading system is the use of adaptive optimization, essentially constantly retuning the parameters of our dynamic model each time new data is encountered in our training, testing and real-time applications. The parameters $\{\mu, \sigma\}$ are constantly tuned using a quasi-local simplex code [27,28] included with the ASA (Adaptive Simulated Annealing) code [12].

We have tested several quasi-local codes for this kind of trading problem, versus using robust ASA adaptive optimizations, and the faster quasi-local codes seem to work quite well for adaptive updates after a zeroth order parameters set is found by ASA [29,30].

# 3. STATISTICAL MECHANICS OF FINANCIAL MARKETS (SMFM)

## 3.1. Statistical Mechanics of Large Systems

Aggregation problems in nonlinear nonequilibrium systems typically are "solved" (accommodated) by having new entities/languages developed at these disparate scales in order to efficiently pass information back and forth between scales. This is quite different from the nature of quasi-equilibrium quasi-linear systems, where thermodynamic or cybernetic approaches are possible. These thermodynamic approaches typically fail for nonequilibrium nonlinear systems.

Many systems are aptly modeled in terms of multivariate differential rate-equations, known as Langevin equations [31],

$$\dot{M}^G = f^G + \hat{g}_j^G \eta^j \,, (G = 1, \cdots, \Lambda) \,, (j = 1, \cdots, N) \,,$$

$$\dot{M}^G = dM^G/dt \,,$$

$$< \eta^j(t) >_\eta = 0 \,, \ < \eta^j(t), \eta^{j'}(t') >_\eta = \delta^{jj'} \delta(t - t') \,, \tag{2}$$

where $f^G$ and $\hat{g}_j^G$ are generally nonlinear functions of mesoscopic order parameters $M^G$, $j$ is a microscopic index indicating the source of fluctuations, and $N \geq \Lambda$. The Einstein convention of summing over repeated indices is used. Vertical bars on an index, e.g., |j|, imply no sum is to be taken on repeated indices.

Via a somewhat lengthy, albeit instructive calculation, outlined in several other papers [11,32,33], involving an intermediate derivation of a corresponding Fokker-Planck or Schrödinger-type equation for the conditional probability distribution $P[M(t)|M(t_0)]$, the Langevin rate Eq. (2) is developed into the more useful probability distribution for $M^G$ at long-time macroscopic time event $t = (u + 1)\theta + t_0$, in terms of a Stratonovich path-integral over mesoscopic Gaussian conditional probabilities [34-38]. Here, macroscopic variables are defined as the long-time limit of the evolving mesoscopic system.

The corresponding Schrödinger-type equation is [36,37]

$$\partial P/\partial t = \frac{1}{2}(g^{GG'}P)_{,GG'} - (g^G P)_{,G} + V \,,$$

$$g^{GG'} = k_T \delta^{jk} \hat{g}_j^G \hat{g}_k^{G'} \,,$$

$$g^G = f^G + \frac{1}{2} \delta^{jk} \hat{g}_j^{G'} \hat{g}_{k,G'}^G \ ,$$

$$[\cdots]_{,G} = \partial[\cdots]/\partial M^G \ . \tag{3}$$

This is properly referred to as a Fokker-Planck equation when $V \equiv 0$. Note that although the partial differential Eq. (3) contains information regarding $M^G$ as in the stochastic differential Eq. (2), all references to $j$ have been properly averaged over. I.e., $\hat{g}_j^G$ in Eq. (2) is an entity with parameters in both microscopic and mesoscopic spaces, but $M$ is a purely mesoscopic variable, and this is more clearly reflected in Eq. (3).

The path integral representation is given in terms of the "Feynman" Lagrangian $L$.

$$P[M_t|M_{t_0}]dM(t) = \int \cdots \int \underset{\sim}{D} M \exp(-S)\delta[M(t_0) = M_0]\delta[M(t) = M_t] \ ,$$

$$S = k_T^{-1} \min \int_{t_0}^{t} dt' L \ ,$$

$$\underset{\sim}{D} M = \lim_{u \to \infty} \prod_{\nu=1}^{u+1} g^{1/2} \prod_G (2\pi\theta)^{-1/2} dM_\nu^G \ ,$$

$$L(\dot{M}^G, M^G, t) = \frac{1}{2} (\dot{M}^G - h^G)g_{GG'}(\dot{M}^{G'} - h^{G'}) + \frac{1}{2} h^G_{;G} + R/6 - V \ ,$$

$$h^G = g^G - \frac{1}{2} g^{-1/2}(g^{1/2} g^{GG'})_{,G'} \ ,$$

$$g_{GG'} = (g^{GG'})^{-1} \ ,$$

$$g = \det(g_{GG'}) \ ,$$

$$h^G_{;G} = h^G_{,G} + \Gamma^F_{GF} h^G = g^{-1/2}(g^{1/2} h^G)_{,G} \ ,$$

$$\Gamma^F_{JK} \equiv g^{LF}[JK, L] = g^{LF}(g_{JL,K} + g_{KL,J} - g_{JK,L}) \ ,$$

$$R = g^{JL} R_{JL} = g^{JL} g^{JK} R_{FJKL} \ ,$$

$$R_{FJKL} = \frac{1}{2} (g_{FK,JL} - g_{JK,FL} - g_{FL,JK} + g_{JL,FK}) + g_{MN}(\Gamma^M_{FK}\Gamma^N_{JL} - \Gamma^M_{FL}\Gamma^N_{JK}) \ . \tag{4}$$

Mesoscopic variables have been defined as $M^G$ in the Langevin and Fokker-Planck representations, in terms of their development from the microscopic system labeled by $j$. The Riemannian curvature term $R$ arises from nonlinear $g_{GG'}$, which is a bona fide metric of this space [36]. Even if a stationary solution, i.e., $\dot{M}^G = 0$, is ultimately sought, a necessarily prior stochastic treatment of $\dot{M}^G$ terms gives rise to these Riemannian "corrections." Even for a constant metric, the term $h^G_{;G}$ contributes to $L$ for a nonlinear mean $h^G$. $V$ may include terms such as $\sum_{T'} J_{T'G} M^G$, where the Lagrange multipliers $J_{T'G}$ are constraints on $M^G$, which are advantageously modeled as extrinsic sources in this representation; they too may be time-dependent.

For our purposes, the above Feynman Lagrangian defines a kernel of the short-time conditional probability distribution, in the curved space defined by the metric, in the limit of continuous time, whose iteration yields the solution of the previous partial differential equation Eq. (3). This differs from the Lagrangian which satisfies the requirement that the action is stationary to the first order in $dt$ — the WKBJ approximation, but which does not include the first-order correction to the WKBJ approximation as does the Feynman Lagrangian. This latter Lagrangian differs from the Feynman Lagrangian, essentially by replacing $R/6$ above by $R/12$ [39]. In this sense, the WKBJ Lagrangian is more useful for some theoretical discussions [40]. However, the use of the Feynman Lagrangian coincides with the numerical method we use for long-time development of our distributions using our PATHINT code for other financial products, e.g., options [4]. This also is consistent with our use of relatively short-time "forecast" of data points using the most probable path [41]

$$dM^G/dt = g^G - g^{1/2}(g^{-1/2} g^{GG'})_{,G'} .$$  (5)

Using the variational principle, $J_{TG}$ may also be used to constrain $M^G$ to regions where they are empirically bound. More complicated constraints may be affixed to $L$ using methods of optimal control theory [42]. With respect to a steady state $\bar{P}$, when it exists, the information gain in state $P$ is defined by

$$\Upsilon[P] = \int \cdots \int DM' \, P \ln (P/\bar{P}) ,$$

$$DM' = DM/dM_{u+1} .$$  (6)

In the economics literature, there appears to be sentiment to define Eq. (2) by the Itô, rather than the Stratonovich prescription. It is true that Itô integrals have Martingale properties not possessed by Stratonovich integrals [43] which leads to risk-neural theorems for markets [44,45], but the nature of the proper mathematics — actually a simple transformation between these two discretizations — should eventually be determined by proper aggregation of relatively microscopic models of markets. It should be noted that virtually all investigations of other physical systems, which are also continuous time models of discrete processes, conclude that the Stratonovich interpretation coincides with reality, when multiplicative noise with zero correlation time, modeled in terms of white noise $\eta^j$, is properly considered as the limit of real noise with finite correlation time [46]. The path integral succinctly demonstrates the difference between the two: The Itô prescription corresponds to the prepoint discretization of $L$, wherein $\theta \dot{M}(t) \rightarrow M_{\nu+1} - M_\nu$ and $M(t) \rightarrow M_\nu$. The Stratonovich prescription corresponds to the midpoint discretization of $L$, wherein $\theta \dot{M}(t) \rightarrow M_{\nu+1} - M_\nu$ and $M(t) \rightarrow \frac{1}{2}(M_{\nu+1} + M_\nu)$. In terms of the functions appearing in the Fokker-Planck Eq. (3), the Itô prescription of the prepoint discretized Lagrangian, $L_I$, is relatively simple, albeit deceptively so because of its nonstandard calculus.

$$L_I(\dot{M}^G, M^G, t) = \frac{1}{2}(\dot{M}^G - g^G)g_{GG'}(\dot{M}^{G'} - g^{G'}) - V \ . \tag{7}$$

In the absence of a nonphenomenological microscopic theory, the difference between a Itô prescription and a Stratonovich prescription is simply a transformed drift [39].

There are several other advantages to Eq. (4) over Eq. (2). Extrema and most probable states of $M^G$, $\ll M^G \gg$, are simply derived by a variational principle, similar to conditions sought in previous studies [47]. In the Stratonovich prescription, necessary, albeit not sufficient, conditions are given by

$$\delta_G L = L_{,G} - L_{,\dot{G}:t} = 0 \ ,$$

$$L_{,\dot{G}:t} = L_{,\dot{G}G'}\dot{M}^{G'} + L_{,\dot{G}\dot{G}'}\ddot{M}^{G'} \ . \tag{8}$$

For stationary states, $\dot{M}^G = 0$, and $\partial \bar{L}/\partial \bar{M}^G = 0$ defines $\ll \bar{M}^G \gg$, where the bars identify stationary variables; in this case, the macroscopic variables are equal to their mesoscopic counterparts. Note that $\bar{L}$ is *not* the stationary solution of the system, e.g., to Eq. (3) with $\partial P/\partial t = 0$. However, in some cases [48], $\bar{L}$ is a definite aid to finding such stationary states. Many times only properties of stationary states are examined, but here a temporal dependence is included. E.g., the $\dot{M}^G$ terms in $L$ permit steady states and their fluctuations to be investigated in a nonequilibrium context. Note that Eq. (8) must be derived from the path integral, Eq. (4), which is at least one reason to justify its development.

## 3.2. Algebraic Complexity Yields Simple Intuitive Results

It must be emphasized that the output of this formalism is not confined to complex algebraic forms or tables of numbers. Because $L$ possesses a variational principle, sets of contour graphs, at different long-time epochs of the path-integral of $P$ over its variables at all intermediate times, give a visually intuitive and accurate decision-aid to view the dynamic evolution of the scenario. For example, this Lagrangian approach permits a quantitative assessment of concepts usually only loosely defined.

$$\text{``Momentum''} = \Pi^G = \frac{\partial L}{\partial(\partial M^G/\partial t)} \ ,$$

$$\text{``Mass''} = g_{GG'} = \frac{\partial^2 L}{\partial(\partial M^G/\partial t)\partial(\partial M^{G'}/\partial t)} \ ,$$

$$\text{``Force''} = \frac{\partial L}{\partial M^G} \ ,$$

$$\text{``}F = ma\text{''}: \ \delta L = 0 = \frac{\partial L}{\partial M^G} - \frac{\partial}{\partial t}\frac{\partial L}{\partial(\partial M^G/\partial t)} \ , \tag{9}$$

where $M^G$ are the variables and $L$ is the Lagrangian. These physical entities provide another form of intuitive, but quantitatively precise, presentation of these analyses. For example, daily newspapers use some of this terminology to discuss the movement of security prices. In this paper, the $\Pi^G$ serve as canonical momenta indicators (CMI) for these systems.

### 3.2.1. Derived Canonical Momenta Indicators (CMI)

The extreme sensitivity of the CMI gives rapid feedback on changes in trends as well as the volatility of markets, and therefore are good indicators to use for trading rules [29]. A time-locked moving average provides manageable indicators for trading signals. This current project uses such CMI developed as a byproduct of the ASA fits described below.

### 3.3. Correlations

In this paper we report results of our one-variable trading model. However, it is straightforward to include multi-variable trading models in our approach, and we have done this, for example, with coupled cash and futures S&P markets.

Correlations between variables are modeled explicitly in the Lagrangian as a parameter usually designated $\rho$. This section uses a simple two-factor model to develop the correspondence between the correlation $\rho$ in the Lagrangian and that among the commonly written Wiener distribution $dz$.

Consider coupled stochastic differential equations for futures $F$ and cash $C$:

$$dF = f^F(F,C)dt + \hat{g}^F(F,C)\sigma_F dz_F \ ,$$

$$dC = f^C(F,C)dt + \hat{g}^C(F,C)\sigma_C dz_C \ ,$$

$$< dz_i >= 0 \ , \ i = \{F,C\} \ ,$$

$$< dz_i(t)dz_j(t') >= dt\delta(t-t') \ , \ i = j \ ,$$

$$< dz_i(t)dz_j(t') >= \rho dt\delta(t-t') \ , \ i \neq j \ ,$$

$$\delta(t-t') = \begin{cases} 0 \, , & t \neq t' \ , \\ 1 \, , & t = t' \ , \end{cases} \tag{10}$$

where $< \, . \, >$ denotes expectations with respect to the multivariate distribution.

These can be rewritten as Langevin equations (in the Itô prepoint discretization)

$$dF/dt = f^F + \hat{g}^F \sigma_F(\gamma^+ \eta_1 + \operatorname{sgn}\rho \ \gamma^- \eta_2) \ ,$$

$$dC/dt = g^C + \hat{g}^C \sigma_C(\operatorname{sgn}\rho \ \gamma^- \eta_1 + \gamma^+ \eta_2) \ ,$$

$$\gamma^{\pm} = \frac{1}{\sqrt{2}}[1 \pm (1-\rho^2)^{1/2}]^{1/2} \ ,$$

$$n_i = (dt)^{1/2}p_i \ , \tag{11}$$

where $p_1$ and $p_2$ are independent [0,1] Gaussian distributions.

The equivalent short-time probability distribution, $P$, for the above set of equations is

$$P = g^{1/2}(2\pi dt)^{-1/2}\exp(-Ldt) \ ,$$

$$L = \frac{1}{2}M^{\dagger}\underline{g}M \ ,$$

$$M = \begin{pmatrix} dF/dt - f^F \\ dC/dt - f^C \end{pmatrix},$$

$$g = \det(\underline{g}) \,. \tag{12}$$

$\underline{g}$, the metric in $\{F, C\}$-space, is the inverse of the covariance matrix,

$$\underline{g}^{-1} = \begin{pmatrix} (\hat{g}^F \sigma_F)^2 & \rho \hat{g}^F \hat{g}^C \sigma_F \sigma_C \\ \rho \hat{g}^F \hat{g}^C \sigma_F \sigma_C & (\hat{g}^C \sigma_C)^2 \end{pmatrix}. \tag{13}$$

The CMI indicators are given by the formulas

$$\Pi^F = \frac{(dF/dt - f^F)}{(\hat{g}^F \sigma_F)^2 (1 - \rho^2)} - \frac{\rho(dC/dt - f^C)}{\hat{g}^F \hat{g}^C \sigma_F \sigma_C (1 - \rho^2)} \,,$$

$$\Pi^C = \frac{(dC/dt - f^C)}{(\hat{g}^C \sigma_C)^2 (1 - \rho^2)} - \frac{\rho(dF/dt - f^F)}{\hat{g}^C \hat{g}^F \sigma_C \sigma_F (1 - \rho^2)} \,. \tag{14}$$

### 3.4. ASA Outline

The algorithm Adaptive Simulated Annealing (ASA) fits short-time probability distributions to observed data, using a maximum likelihood technique on the Lagrangian. This algorithm has been developed to fit observed data to a theoretical cost function over a $D$-dimensional parameter space [13], adapting for varying sensitivities of parameters during the fit. The ASA code can be obtained at no charge, via WWW from http://www.ingber.com/ or via FTP from ftp.ingber.com [12].

### 3.4.1. General Description

Simulated annealing (SA) was developed in 1983 to deal with highly nonlinear problems [49], as an extension of a Monte-Carlo importance-sampling technique developed in 1953 for chemical physics problems. It helps to visualize the problems presented by such complex systems as a geographical terrain. For example, consider a mountain range, with two "parameters," e.g., along the North–South and East–West directions. We wish to find the lowest valley in this terrain. SA approaches this problem similar to using a bouncing ball that can bounce over mountains from valley to valley. We start at a high "temperature," where the temperature is an SA parameter that mimics the effect of a fast moving particle in a hot object like a hot molten metal, thereby permitting the ball to make very high bounces and being able to bounce over any mountain to access any valley, given enough bounces. As the temperature is made relatively colder, the ball cannot bounce so high, and it also can settle to become trapped in relatively smaller ranges of valleys.

We imagine that our mountain range is aptly described by a "cost function." We define probability distributions of the two directional parameters, called generating distributions since they generate possible valleys or states we are to explore. We define another distribution, called the acceptance distribution, which depends on the difference of cost functions of the present generated valley we are to explore and the last saved lowest valley. The acceptance distribution decides probabilistically whether to stay in a new lower valley or to bounce out of it. All the generating and acceptance distributions depend on temperatures.

In 1984 [50], it was established that SA possessed a proof that, by carefully controlling the rates of cooling of temperatures, it could statistically find the best minimum, e.g., the lowest valley of our example above. This was good news for people trying to solve hard problems which could not be solved by other algorithms. The bad news was that the guarantee was only good if they were willing to run SA forever. In 1987, a method of fast annealing (FA) was developed [51], which permitted lowering the temperature exponentially faster, thereby statistically guaranteeing that the minimum could be found in some finite time. However, that time still could be quite long. Shortly thereafter, Very Fast Simulated Reannealing (VFSR) was developed in 1987 [13], now called Adaptive Simulated Annealing (ASA), which is exponentially faster than FA.

ASA has been applied to many problems by many people in many disciplines [15,16,52]. The feedback of many users regularly scrutinizing the source code ensures its soundness as it becomes more flexible and powerful.

### 3.4.2. Mathematical Outline

ASA considers a parameter $\alpha_k^i$ in dimension $i$ generated at annealing-time $k$ with the range

$$\alpha_k^i \in [A_i, B_i] \,, \tag{15}$$

calculated with the random variable $y^i$,

$$\alpha_{k+1}^i = \alpha_k^i + y^i (B_i - A_i) \,,$$

$$y^i \in [-1, 1] \ .$$                                                           (16)

The generating function $g_T(y)$ is defined,

$$g_T(y) = \prod_{i=1}^{D} \frac{1}{2(|y^i| + T_i)\ln(1 + 1/T_i)} \equiv \prod_{i=1}^{D} g_T^i(y^i) \ ,$$   (17)

where the subscript $i$ on $T_i$ specifies the parameter index, and the $k$-dependence in $T_i(k)$ for the annealing schedule has been dropped for brevity. Its cumulative probability distribution is

$$G_T(y) = \int_{-1}^{y^1} \cdots \int_{-1}^{y^D} dy'^1 \cdots dy'^D \, g_T(y') \equiv \prod_{i=1}^{D} G_T^i(y^i) \ ,$$

$$G_T^i(y^i) = \frac{1}{2} + \frac{\operatorname{sgn}(y^i)}{2} \frac{\ln(1 + |y^i|/T_i)}{\ln(1 + 1/T_i)} \ .$$     (18)

$y^i$ is generated from a $u^i$ from the uniform distribution

$$u^i \in U[0, 1] \ ,$$

$$y^i = \operatorname{sgn}(u^i - \frac{1}{2})T_i[(1 + 1/T_i)^{|2u^i - 1|} - 1] \ .$$       (19)

It is straightforward to calculate that for an annealing schedule for $T_i$

$$T_i(k) = T_{0i} \exp(-c_i k^{1/D}) \ ,$$                                          (20)

a global minima statistically can be obtained. I.e.,

$$\sum_{k_0}^{\infty} g_k \approx \sum_{k_0}^{\infty} [\prod_{i=1}^{D} \frac{1}{2|y^i|c_i}] \frac{1}{k} = \infty \ .$$   (21)

Control can be taken over $c_i$, such that

$$T_{fi} = T_{0i} \exp(-m_i) \text{ when } k_f = \exp n_i \ ,$$

$$c_i = m_i \exp(-n_i/D) \ ,$$                                                      (22)

where $m_i$ and $n_i$ can be considered "free" parameters to help tune ASA for specific problems.

ASA has over 100 OPTIONS available for tuning. A few important ones were used in this project.

### 3.4.3. Reannealing

Whenever doing a multi-dimensional search in the course of a complex nonlinear physical problem, inevitably one must deal with different changing sensitivities of the $\alpha^i$ in the search. At any given annealing-time, the range over which the relatively insensitive parameters are being searched can be "stretched out" relative to the ranges of the more sensitive parameters. This can be accomplished by periodically rescaling the annealing-time $k$, essentially reannealing, every hundred or so acceptance-events (or at some user-defined modulus of the number of accepted or generated states), in terms of the sensitivities $s_i$ calculated at the most current minimum value of the cost function, $C$,

$$s_i = \partial C / \partial \alpha^i \ . \tag{23}$$

In terms of the largest $s_i = s_{\max}$, a default rescaling is performed for each $k_i$ of each parameter dimension, whereby a new index $k'_i$ is calculated from each $k_i$,

$$k_i \rightarrow k'_i \ ,$$

$$T'_{ik'} = T_{ik}(s_{\max}/s_i) \ ,$$

$$k'_i = (\ln(T_{i0}/T_{ik'})/c_i)^D \ . \tag{24}$$

$T_{i0}$ is set to unity to begin the search, which is ample to span each parameter dimension.

### 3.4.4. Quenching

Another adaptive feature of ASA is its ability to perform quenching in a methodical fashion. This is applied by noting that the temperature schedule above can be redefined as

$$T_i(k_i) = T_{0i} \exp(-c_i k_i^{Q_i/D}) \ ,$$

$$c_i = m_i \exp(-n_i Q_i/D) \ , \tag{25}$$

in terms of the "quenching factor" $Q_i$. The sampling proof fails if $Q_i > 1$ as

$$\sum_k \prod^D 1/k^{Q_i/D} = \sum_k 1/k^{Q_i} < \infty \ . \tag{26}$$

This simple calculation shows how the "curse of dimensionality" arises, and also gives a possible way of living with this disease. In ASA, the influence of large dimensions becomes clearly focussed on the exponential of the power of $k$ being $1/D$, as the annealing required to properly sample the space becomes prohibitively slow. So, if resources cannot be committed to properly sample the space, then for some systems perhaps the next best procedure may be to turn on quenching, whereby $Q_i$ can become on the order of the size of number of dimensions.

The scale of the power of $1/D$ temperature schedule used for the acceptance function can be altered in a similar fashion. However, this does not affect the annealing proof of ASA, and so this may used without damaging the sampling property.

### 3.4.5. Avoiding Repeating Cost Functions

Doing a recursive optimization is very CPU expensive, as essentially the cross-product of parameter spaces among the various levels of optimization is required.

Therefore, we have used an ASA OPTION for some of the parameters in the outer-shell trading model optimization of training sets, ASA_QUEUE, which sets up a first-in first-out (FIFO) queue, of user-defined size Queue_Size to collect generated states. When a new state is generated, its parameters are tested, within specified resolutions of a user-defined array Queue_Resolution[]. When parameters sets are repeated within this queue, the saved value of the cost function is returned without having to repeat the calculation.

### 3.4.6. Multiple Local Minima

Our criteria for the global minimum of our cost function is minus the largest profit over a selected training data set (or in some cases, this value divided by the maximum drawdown). However, in many cases this may not give us the best set of parameters to find profitable trading in test sets or in real-time trading. Other considerations such as the total number of trades developed by the global minimum versus other close local minima may be relevant. For example, if the global minimum has just a few trades, while some nearby local minima (in terms of the value of the cost function) have many trades and was profitable in spite of our slippage factors, then the scenario with more trades might be more statistically dependable to deliver profits across testing and real-time data sets.

Therefore, for the outer-shell global optimization of training sets, we have used an ASA OPTION, MULTI_MIN, which saves a user-defined number of closest local minima within a user-defined resolution of the parameters. We then examine these results under several testing sets.

## 4. TRADING SYSTEM

### 4.1. Use of CMI

As the CMI formalism carries the relevant information regarding the prices dynamics, we have used it as a signal generator for an automated trading system for S&P futures.

Based on a previous work [30] applied to daily closing data, the overall structure of the trading system consists in 2 layers, as follows: We first construct the "short-time" Lagrangian function in the Itô representation (with the notation introduced in Section 3.3)

$$L(i|i-1) = \frac{1}{2\sigma^2 F_{i-1}^{2x}} \left( \frac{dF_i}{dt} - f^F \right)^2 \tag{27}$$

with $i$ the post-point index, corresponding to the one factor price model

$$dF = f^F dt + \sigma F^x dz(t) , \tag{28}$$

where $f^F$ and $\sigma > 0$ are taken to be constants, $F(t)$ is the S&P future price, and $dz$ is the standard Gaussian noise with zero mean and unit standard deviation. We perform a global, maximum likelihood fit to the whole set of price data using ASA. This procedure produces the optimization parameters $\{x, f^F\}$ that are used to generate the CMI. One computational approach was to fix the diffusion multiplier $\sigma$ to 1 during training for convenience, but used as free parameters in the adaptive testing and real-time fits. Another approach was to fix the scale of the volatility, using an improved model,

$$dF = f^F dt + \sigma \left( \frac{F}{<F>} \right)^x dz(t) , \tag{29}$$

where $\sigma$ now is calculated as the standard deviation of the price increments $\Delta F/dt^{1/2}$, and $<F>$ is just the average of the prices.

As already remarked, to enhance the CMI sensitivity and response time to local variations (across a certain window size) in the distribution of price increments, the momenta are generated applying an adaptive procedure, i.e., after each new data reading another set of $\{f^F, \sigma\}$ parameters are calculated for the last window of data, with the exponent $x$ — a contextual indicator of the noise statistics — fixed to the value obtained from the global fit.

The CMI computed in this manner are fed into the outer shell of the trading system, where an AI-type optimization of the trading rules is executed, using ASA once again.

The trading rules are a collection of logical conditions among the CMI, prices and optimization parameters that could be window sizes, time resolutions, or trigger thresholds. Based on the relationships between CMI and optimization parameters, a trading decision is made. The cost function in the outer shell is either the overall equity or the risk-adjusted profit (essentially the return). The inner and outer shell optimizations are coupled through some of the optimization parameters (e.g., time resolution of the data, window sizes), which justifies the recursive nature of the optimization.

Next, we describe in more details the concrete implementation of this system.

## 4.2. Data Processing

The CMI formalism is general and by construction permits us to treat multivariate coupled markets. In certain conditions (e.g., shorter time scales of data), and also due to superior scalability across different markets, it is desirable to have a trading system for a single instrument, in our case the S&P futures contracts that are traded electronically on Chicago Mercantile Exchange (CME). The focus of our system was intra-day trading, at time scales of data used in generating the buy/sell signals from 10 to 60 secs. In particular, we here give some results obtained when using data having a time resolution $\Delta t$ of 55 secs (the time between consecutive data elements is 55 secs). This particular choice of time resolution reflects the set of optimization parameters that have been applied in actual trading.

It is important to remark that a data point in our model does not necessarily mean an actual tick datum. For some trading time scales and for noise reduction purposes, data is pre-processed into sampling bins of length $\Delta t$ using either a standard averaging procedure or spectral filtering (e.g., wavelets, Fourier) of the tick data. Alternatively, the data can be defined in block bins that contain disjoint sets of averaged tick data, or in overlapping bins of widths $\Delta t$ that update at every $\Delta t' < \Delta t$, such that an effective resolution $\Delta t'$ shorter than the width of the sampling bin is obtained. We present here work in which we have used disjoint block bins and a standard average of the tick data with time stamps falling within the bin width.

In Figs. 1 and 2 we present examples of S&P futures data sampled with 55 secs resolution. We remark that there are several time scales — from mins to one hour — at which an automated trading system might extract profits. Fig. 2 illustrates the sustained short trading region of 1.5 hours and several shorter long and short trading regions of about 10-20 mins. Fig. 1 illustrates that the profitable regions are prominent even for data representing a relatively flat market period. I.e., June 20 shows an uptrend region of about 1 hour 20 mins and several short and long trading domains between 10 mins and 20 mins. In both situations, there are a larger number of opportunities at time resolutions smaller than 5 mins.

The time scale at which we sample the data for trading is itself a parameter that is extracted from the optimization of the trading rules and of the Lagrangian cost function Eq. (27). This is one of the coupling parameters between the inner- and the outer-shell optimizations.

### 4.3. Inner-Shell Optimization

A cycle of optimization runs has three parts, training and testing, and finally real-time use — a variant of testing. Training consists in choosing a data set and performing the recursive optimization, which produces optimization parameters for trading. In our case there are six parameters: the time resolution $\Delta t$ of price data, the length of window $W$ used in the local fitting procedures and in computation of moving averages of trading signals, the drift $f^F$, volatility coefficient $\sigma$ and exponent $x$ from Eq. (28), and a multiplicative factor $M$ necessary for the trading rules module, as discussed below.

The optimization parameters computed from the training set are applied then to various test sets and final profit/loss analysis are produced. Based on these, the best set of optimization parameters are chosen to be applied in real-time trading runs. We remark once again that a single training data set could support more than one profitable sets of parameters and can be a function of the trader's interest and the specific market dynamics targeted (e.g., short/long time scales). The optimization parameters corresponding to the global minimum in the training session may not necessarily represent the parameters that led to robust profits across real-time data.

The training optimization occurs in two inter-related stages. An inner-shell maximum likelihood optimization over all training data is performed. The cost function that is fitted to data is the effective action constructed from the Lagrangian Eq. (27) including the pre-factors coming from the measure element in the expression of the short-time probability distribution Eq. (12). This is based on the fact [39] that in the context of Gaussian multiplicative stochastic noise, the macroscopic transition probability $P(F, t|F', t')$ to start with the price $F'$ at $t'$ and reach the price $F$ at $t$ is determined by the short-time Lagrangian Eq. (27),

$$P(F, t|F', t') = \frac{1}{(2\pi\sigma^2 F_{i-1}^{2x} dt_i)^{1/2}} \exp\left(-\sum_{i=1}^{N} L(i|i-1)dt_i\right), \tag{30}$$

with $dt_i = t_i - t_{i-1}$. Recall that the main assumption of our model is that price increments (or the logarithm of price ratios, depending on which variables are considered independent) could be described by a system of coupled stochastic, non-linear equations as in Eq. (10). These equations are deceptively simple in structure, yet depending on the functional form of the drift coefficients and the multiplicative noise, they could describe a variety of interactions between financial instruments in various market conditions (e.g., constant elasticity of variance model [53], stochastic volatility models, etc.). In particular, this type of models include the case of Black-Scholes price dynamics ($x = 1$).

In the system presented here, we have applied the model from Eq. (28). The fitted parameters were the drift coefficient $f^F$ and the exponent $x$. In the case of a coupled futures and cash system, besides the corresponding values of $f^F$ and $x$ for the cash index, another parameter, the correlation coefficient $\rho$ as introduced in Eq. (10), must be considered.

### 4.4. Trading Rules (Outer-Shell) Recursive Optimization

In the second part of the training optimization, we calculate the CMI and execute trades as required by a selected set of trading rules based on CMI values, price data or combinations of both indicators.

Recall that three external shell optimization parameters are defined: the time resolution $\Delta t$ of the data expressed as the time interval between consecutive data points, the window length $W$ (in number of time epochs or data points) used in the adaptive calculation of CMI, and a numerical coefficient $M$ that scales the momentum uncertainty discussed below.

At each moment a local refit of $f^F$ and $\sigma$ over data in the local window $W$ is executed, moving the window $M$ across the training data set and using the zeroth order optimization parameters $f^F$ and $x$ resulting from the inner-shell optimization as a first guess. It was found that a faster quasi-local code is sufficient for computational purposes for these adaptive updates. In more complicated models, ASA can be successfully applied recursively, although in real-time trading the response time of the system is a major factor that requires attention.

All expressions that follow can be generalized to coupled systems in the manner described in Section 3. Here we use the one factor nonlinear model given by Eq. (28). At each time epoch we calculate the following momentum related quantities:

$$\Pi^F = \frac{1}{\sigma^2 F^{2x}} \left( \frac{dF}{dt} - f^F \right),$$

$$\Pi_0^F = -\frac{f^F}{\sigma^2 F^{2x}},$$

$$\Delta \Pi^F = < (\Pi^F - < \Pi^F >)^2 >^{1/2} = \frac{1}{\sigma F^x \sqrt{dt}},$$

(31)

where we have used $< \Pi^F >= 0$ as implied by Eqs. (28) and (27). In the previous expressions, $\Pi^F$ is the CMI, $\Pi_0^F$ is the neutral line or the momentum of a zero change in prices, and $\Delta \Pi^F$ is the uncertainty of momentum. The last quantity reflects the Heisenberg principle, as derived from Eq. (28) by calculating

$$\Delta F \equiv < (dF - < dF >)^2 >^{1/2} = \sigma F^x \sqrt{dt},$$

$$\Delta \Pi^F \Delta F \geq 1,$$

(32)

where all expectations are in terms of the exact noise distribution, and the calculation implies the Itô approximation (equivalent to considering non-anticipative functions). Various moving averages of these momentum signals are also constructed. Other dynamical quantities, as the Hamiltonian, could be used as well. (By analogy to the energy concept, we found that the Hamiltonian carries information regarding the overall trend of the market, giving another useful measure of price volatility.)

Regarding the practical implementation of the previous relations for trading, some comments are necessary. In terms of discretization, if the CMI are calculated at epoch $i$, then $dF_i = F_i - F_{i-1}$, $dt_i = t_i - t_{i-1} = \Delta t$, and all prefactors are computed at moment $i-1$ by the Itô prescription (e.g., $\sigma F^x = \sigma F_{i-1}^x$). The momentum uncertainty band $\Delta \Pi^F$ can be calculated from the discretized theoretical value Eq. (31), or by computing the estimator of the standard deviation from the actual time series of $\Pi^F$.

There are also two ways of calculating averages over CMI values: One way is to use the set of local optimization parameters $\{f^F, \sigma\}$ obtained from the local fit procedure in the current window $W$ for all CMI data within that window (local-model average). The second way is to calculate each CMI in the current local window $W$ with another set $\{f^F, \sigma\}$ obtained from a previous local fit window measured from the CMI data backwards $W$ points (multiple-models averaged, as each CMI corresponds to a different model in terms of the fitting parameters $\{f^F, \sigma\}$).

The last observation is that the neutral line divides all CMI in two classes: long signals, when $\Pi^F > \Pi_0^F$, as any CMI satisfying this condition indicates a positive price change, and short signals when $\Pi^F < \Pi_0^F$, which reflects a negative price change.

After the CMI are calculated, based on their meaning as statistical momentum indicators, trades are executed following a relatively simple model: Entry in and exit from a long (short) trade points are defined as points where the value of CMIs is greater (smaller) than a certain fraction of the uncertainty band $M \Delta \Pi^F$ ($-M \Delta \Pi^F$), where $M$ is the multiplicative factor mentioned in the beginning of this subsection. This is a choice of a symmetric trading rule, as $M$ is the same for long and short trading signals, which is suitable for volatile markets without a sustained trend, yet without diminishing too severely profits in a strictly bull or bear region.

Inside the momentum uncertainty band, one could define rules to stay in a previously open trade, or exit immediately, because by its nature the momentum uncertainty band implies that the probabilities of price movements in either direction (up or down) are balanced. From another perspective, this type of trading rule exploits the relaxation time of a strong market advance or decline, until a trend reversal occurs or it becomes more probable.

Other sets of trading rules are certainly possible, by utilizing not only the current values of the momenta indicators, but also their local-model or multiple-models averages. A trading rule based on the maximum distance between the current CMI data $\Pi_i^F$ and the neutral line $\Pi_0^F$ shows faster response to markets evolution and may be more suitable to automatic trading in certain conditions.

Stepping through the trading decisions each trading day of the training set determined the profit/loss of the training set as a single value of the outer-sell cost function. As ASA importance-sampled the outer-shell parameter space $\{\Delta t, W, M\}$, these parameters are fed into the inner shell, and a new inner-shell recursive optimization cycle begins. The final values for the optimization parameters in the training set are fixed when the largest net profit (calculated from the total equity by subtracting the transactions costs defined by the slippage factor) is realized. In practice, we have collected optimization parameters from multiple local minima that are near the global minimum (the outer-shell cost function is defined with the sign reversed) of the training set.

The values of the optimization parameters $\{\Delta t, W, M, f^F, \sigma, x\}$ resulting from a training cycle are then applied to out-of-sample test sets. During the test run, the drift coefficient $f^F$ and the volatility coefficient $\sigma$ are refitted adaptively as described previously. All other parameters are fixed. We have mentioned that the optimization parameters corresponding to the highest profit in the training set may not be the sufficiently robust across test sets. Then, for all test sets, we have tested optimization parameters related to the multiple minima (i.e., the global maximum profit, the second best profit, etc.) resulting from the training set.

We performed a bootstrap-type reversal of the training-test sets (repeating the training runs procedures using one of the test sets, including the previous training set in the new batch of test sets), followed by a selection of the best parameters across all data sets. This is necessary to increase the chances of successful trading sessions in real-time.

## 5. RESULTS

### 5.1. Alternative Algorithms

In the previous sections we noted that there are different combinations of methods of processing data, methods of computing the CMI and various sets of trading rules that need to be tested — at least in a sampling manner — before launching trading runs in real-time:

1. Data can be preprocessed in block or overlapping bins, or forecasted data derived from the most probable transition path [41] could be used as in one of our most recent models.

2. Exponential smoothing, wavelets or Fourier decomposition can be applied for statistical processing. We presently favor exponential moving averages.

3. The CMI can be calculated using averaged data or directly with tick data, although the optimization parameters were fitted from preprocessed (averaged) price data.

4. The trading rules can be based on current signals (no average is performed over the signal themselves), on various averages of the CMI trading signals, on various combination of CMI data (momenta, neutral line, uncertainty band), on symmetric or asymmetric trading rules, or on mixed price-CMI trading signals.

5.  Different models (one and two-factors coupled) can be applied to the same market instrument, e.g., to define complementary indicators.

The selection process evidently must consider many specific economic factors (e.g., liquidity of a given market), besides all other physical, mathematical and technical considerations.  In the work presented here, as we tested our system and using previous experience, we focused toward S&P500 futures electronic trading, using block processed data, and symmetric, local-model and multiple-models trading rules.  In Table 1 we show results obtained for several training and testing sets in the mentioned context.

## 5.2.  Trading System Design

The design of a successful electronic trading system is complex as it must incorporate several aspects of a trader's actions that sometimes are difficult to translate into computer code.  Three important features that must be implemented are factoring in the transactions costs, devising money management techniques, and coping with execution deficiencies.

Generally, most trading costs can be included under the "slippage factor," although this could easily lead to poor estimates.  Given that the margin of profits from exploiting market inefficiencies are thin, a high slippage factor can easily result in a non-profitable trading system.  In our situation, for testing purposes we used a $35 slippage factor per buy & sell order, a value we believe is rather high for an electronic trading environment, although it represents less than three ticks of a mini-S&P futures contract. (The mini-S&P is the S&P futures contract that is traded electronically on CME.)  This higher value was chosen to protect ourselves against the bid-ask spread, as our trigger price (at what price the CMI was generated) and execution price (at what price a trade signaled by a CMI was executed) were taken to be equal to the trading price.  (We have changed this aspect of our algorithm in later models.)  The slippage is also strongly influenced by the time resolution of the data.  Although the slippage is linked to bid-ask spreads and markets volatility in various formulas [54], the best estimate is obtained from experience and actual trading.

Money management was introduced in terms of a trailing stop condition that is a function of the price volatility, and a stop-loss threshold that we fixed by experiment to a multiple of the mini-S&P contract value ($200).  It is tempting to tighten the trailing stop or to work with a small stop-loss value,

yet we found — as otherwise expected — that higher losses occurred as the signals generated by our stochastic model were bypassed.

Regarding the execution process, we have to account for the response of the system to various execution conditions in the interaction with the electronic exchange: partial fills, rejections, uptick rule (for equity trading), etc. Except for some special conditions, all these steps must be automated.

### 5.3. Some Explicit Results

Typical CMI data in Figs. 3 and 4 (obtained from real-time trading after a full cycle of training-testing was performed) are related to the price data in Figs. 1 and 2. We have plotted the fastest (55 secs apart) CMI values $\Pi^F$, the neutral line $\Pi_0^F$ and the uncertainty band $\Delta\Pi^F$. All CMI data were produced using the optimization parameters set $\{55 \text{ secs}, 88 \text{ epochs}, 0.15\}$ of the second-best net profit obtained with the training set "4D ESM0 0321-0324" (Table 1).

Although the CMIs exhibit an inherently ragged nature and oscillate around a zero mean value within the uncertainty band — the width of which is decreasing with increasing price volatility, as the uncertainty principle would also indicate — time scales at which the CMI average or some persistence time are not balanced about the neutral line.

These characteristics, which we try to exploit in our system, are better depicted in Figs. 5 and 6. One set of trading signals, the local-model average of the neutral line $< \Pi_0^F >$ and the uncertainty band multiplied by the optimization factor $M = 0.15$, and centered around the theoretical zero mean of the CMI, is represented versus time. Note entry points in a short trading position ($< \Pi_0^F > \, > \, M \, \Delta\Pi^F$) at around 10:41 (Fig. 5 in conjunction with S&P data in Fig. 1) with a possible exit at 11:21 (or later), and a first long entry ($< \Pi_0^F > \, < \, - M \, \Delta\Pi^F$) at 12:15. After 14:35, a stay long region appears ($< \Pi_0^F > \, < \, 0$), which indicates correctly the price movement in Fig.1.

In Fig. 6 corresponding to June 22 price data from Fig. 2, a first long signal is generated at around 12:56 and a first short signal is generated at 14:16 that reflects the long downtrend region in Fig. 2. Due to the averaging process, a time lag is introduced, reflected by the long signal at 12:56 in Fig. 4, related to a past upward trend seen in Fig. 2; yet the neutral line relaxes rather rapidly (given the 55 sec time resolution and the window of $88 \approx 1.5$ hour) toward the uncertainty band. A judicious choice of trading rules, or avoiding standard averaging methods, helps in controlling this lag problem.

In Tables 1 and 2 we show some results obtained for several training and testing sets following the procedures described at the end of the previous section. In both tables, under the heading "Training" or "Testing Set" we specify the data set used (e.g., "4D ESM0 0321-0324" represents four days of data from the mini-S&P futures contract that expired in June). The type of trading rules used is identified by "LOCAL MODEL" or "MULTIPLE MODELS" tags. These tags refer to how we calculate the averages of the trading signals: either by using a single pair of optimization parameters $\{f^F, \sigma\}$ for all CMI data within the current adaptive fit window, or a different pair $\{f^F, \sigma\}$ for each CMI data. In the "Statistics" column we report the net (subtracting the slippage) profit or loss (in parenthesis) across the whole data set, the total number of trades ("trades"), the number of days with positive balance ("days +"), and the percentage of winning trades ("winners"). The "Parameters" are the optimization parameters resulting from the first three best profit maxima of each listed training set. The parameters are listed in the order $\{\Delta t, W, M\}$, with the data time resolution $\Delta t$ measured in seconds, the length of the local fit window $W$ measured in time epochs, and $M$ the numerical coefficient of the momentum uncertainty band $\Delta \Pi^F$.

Recall that the trading rules presented are symmetric (the long and short entry/exit signals are controlled by the same $M$ factor), and we apply a stay-long condition if the neutral-line is below the average momentum $< \Pi^F >= 0$ and stay-short if $\Pi_0^F > 0$. The drift $f^F$ and volatility coefficient $\sigma$ are refitted adaptively and the exponent $x$ is fixed to the value obtained in the training set. Typical values are $f^F \in \pm [0.003 : 0.05]$, $x \in \pm [0.01 : 0.03]$. During the local fit, due to the shorter time scale involved, the drift may increase by a factor of ten, and $\sigma \in [0.01 : 1.2]$.

Comparing the data in the training and testing tables, we note that the most robust optimization factors — in terms of maximum cumulative profit resulted for all test sets — do not correspond to the maximum profit in the training sets: For the local-model rules, the optimum parameters are $\{55, 88, 0.15\}$, and for the multiple models rules the optimum set is $\{45, 72, 0.2\}$, both realized by the training set "4D ESM0 0321-0324."

Other observations are that, for the data presented here, the multiple-models averages trading rules consistently performed better and are more robust than the local-model averages trading rules. The number of trades is similar, varying between 15 and 35 (eliminating cumulative values smaller than 10 trades), and the time scale of the local fit is rather long in the 30 mins to 1.5 hour range. In the current set-up, this extended time scale implies that is advisable to deploy this system as a trader-assisted tool.

An important factor is the average length of the trades. For the type of rules presented in this work, this length is of several minutes, up to one hour, as the time scale of the local fit window mentioned above suggested.

Related to the length of a trade is the length of a winning long/short trade in comparison to a losing long/short trade. Our experience indicates that a ratio of 2:1 between the length of a winning trade and the length of a losing trade is desirable for a reliable trading system. Here, using the local-model trading rules seems to offer an advantage, although this is not as clear as one would expect.

Finally, the training sets data (Table 1) show that the percentage of winners is markedly higher in the case of multiple-models average than local-average trading rules. In the testing sets (Table 2) the situation is almost reversed, albeit the overall profits (losses) are higher (smaller) in the multiple-model case. Apparently, the multiple-model trading rules can stay in winning trades longer to increase profits, relative to losses incurred with these rules in losing trades. (In the testing sets, this correlates with the higher number of trades executed using local-model trading rules.)

## 6. CONCLUSIONS

### 6.1. Main Features

The main stages of building and testing this system were:

1. We developed a multivariate, nonlinear statistical mechanics model of S&P futures and cash markets, based on a system of coupled stochastic differential equations.

2. We constructed a two-stage, recursive optimization procedure using methods of ASA global optimization: An inner-shell extracts the characteristics of the stochastic price distribution and an outer-shell generates the technical indicators and optimize the trading rules.

3. We trained the system on different sets of data and retained the multiple minima generated (corresponding to the global maximum net profit realized and the neighboring profit maxima).

4. We tested the system on out-of-sample data sets, searching for most robust optimization parameters to be used in real-time trading. Robustness was estimated by the cumulative profit/loss across diverse test sets, and by testing the system against a bootstrap-type reversal of training-testing sets in the optimization cycle.

Modeling the market as a dynamical physical system makes possible a direct representation of empirical notions as market momentum in terms of CMI derived naturally from our theoretical model. We have shown that other physical concepts as the uncertainty principle may lead to quantitative signals (the momentum uncertainty band $\Delta\Pi^F$) that captures other aspects of market dynamics and which can be used in real-time trading.

## 6.2. Summary

We have presented a description of a trading system composed of an outer-shell trading-rule model and an inner-shell nonlinear stochastic dynamic model of the market of interest, S&P500. The inner-shell is developed adhering to the mathematical physics of multivariate nonlinear statistical mechanics, from which we develop indicators for the trading-rule model, i.e., canonical momenta indicators (CMI). We have found that keeping our model faithful to the underlying mathematical physics is not a limiting constraint on profitability of our system; quite the contrary.

An important result of our work is that the ideas for our algorithms, and the proper use of the mathematical physics faithful to these algorithms, must be supplemented by many practical considerations en route to developing a profitable trading system. For example, since there is a subset of parameters, e.g., time resolution parameters, shared by the inner- and outer-shell models, recursive optimization is used to get the best fits to data, as well as developing multiple minima with approximate similar profitability. The multiple minima often have additional features requiring consideration for real-time trading, e.g., more trades per day increasing robustness of the system, etc. The nonlinear stochastic nature of our data required a robust global optimization algorithm. The output of these parameters from these training sets were then applied to testing sets on out-of-sample data. The best models and parameters were then used in real-time by traders, further testing the models as a precursor to eventual deployment in automated electronic trading.

We have used methods of statistical mechanics to develop our inner-shell model of market dynamics and a heuristic AI type model for our outer-shell trading-rule model, but there are many other candidate (quasi-)global algorithms for developing a cost function that can be used to fit parameters to data, e.g., neural nets, fractal scaling models, etc. To perform our fits to data, we selected an algorithm, Adaptive Simulated Annealing (ASA), that we were familiar with, but there are several other candidate algorithms that likely would suffice, e.g., genetic algorithms, tabu search, etc.

We have shown that a minimal set of trading signals (the CMI, the neutral line representing the momentum of the trend of a given time window of data, and the momentum uncertainty band) can generate a rich and robust set of trading rules that identify profitable domains of trading at various time scales. This is a confirmation of the hypothesis that markets are not efficient, as noted in other studies [11,30,55].

### 6.3.  Future Directions

Although this paper focused on trading of a single instrument, the futures S&P 500, the code we have developed can accommodate trading on multiple markets. For example, in the case of tick-resolution coupled cash and futures markets, which was previously prototyped for inter-day trading [29,30], the utility of CMI stems from three directions:

(a) The inner-shell fitting process requires a global optimization of all parameters in both futures and cash markets.

(b) The CMI for futures contain, by our Lagrangian construction, the coupling with the cash market through the off-diagonal correlation terms of the metric tensor. The correlation between the futures and cash markets is explicitly present in all futures variables.

(c) The CMI of both markets can be used as complimentary technical indicators for trading in futures market.

Several near term future directions are of interest: orienting the system toward shorter trading time scales (10-30 secs) more suitable for electronic trading, introducing fast response "averaging" methods and time scale identifiers (exponential smoothing, wavelets decomposition), identifying mini-crashes points using renormalization group techniques, investigating the use of CMI in pattern-recognition based trading rules, and exploring the use of forecasted data evaluated from most probable transition path formalism.

Our efforts indicate the invaluable utility of a joint approach (AI-based and quantitative) in developing automated trading systems.

**6.4.  Standard Disclaimer**

We must emphasize that there are no claims that all results are positive or that the present system is a safe source of riskless profits.  There as many negative results as positive, and a lot of work is necessary to extract meaningful information.  Our purpose here is to describe an approach to developing an electronic trading system complementary to those based on neural-networks type technical analysis and pattern recognition methods.  The system discussed in this paper is rooted in the physical principles of nonequilibrium statistical mechanics, and we have shown that there are conditions under which such a model can be successful.

**ACKNOWLEDGMENTS**

We thank Donald Wilson for his financial support.  We thank K.S. Balasubramaniam and Colleen Chen for their programming support and participation in formulating parts of our trading system.  Data was extracted from the DRW Reuters feed.

**FIGURE CAPTIONS**

Figure 1.  Futures and cash data, contract ESU0 June 20: solid line — futures; dashed line — cash.

Figure 2.  Futures and cash data, contract ESU0 June 22: solid line — futures; dashed line — cash.

Figure 3.  CMI data, real-time trading June 20: solid line — CMI; dashed line — neutral line; dotted line — uncertainty band.

Figure 4.  CMI data, real-time trading, June 22: solid line — CMI; dashed line — neutral line; dotted line — uncertainty band.

Figure 5.  CMI trading signals, real-time trading June 20: dashed line — local-model average of the neutral line; dotted line — uncertainty band multiplied by the optimization parameter $M = 0.15$.

Figure 6.  CMI trading signals, real-time trading June 22: dashed line — local-model average of the neutral line; dotted line — uncertainty band multiplied by the optimization parameter $M = 0.15$.

**TABLE CAPTIONS**

Table 1.  Matrix of Training Runs.

Table 2.  Matrix of Testing Runs.

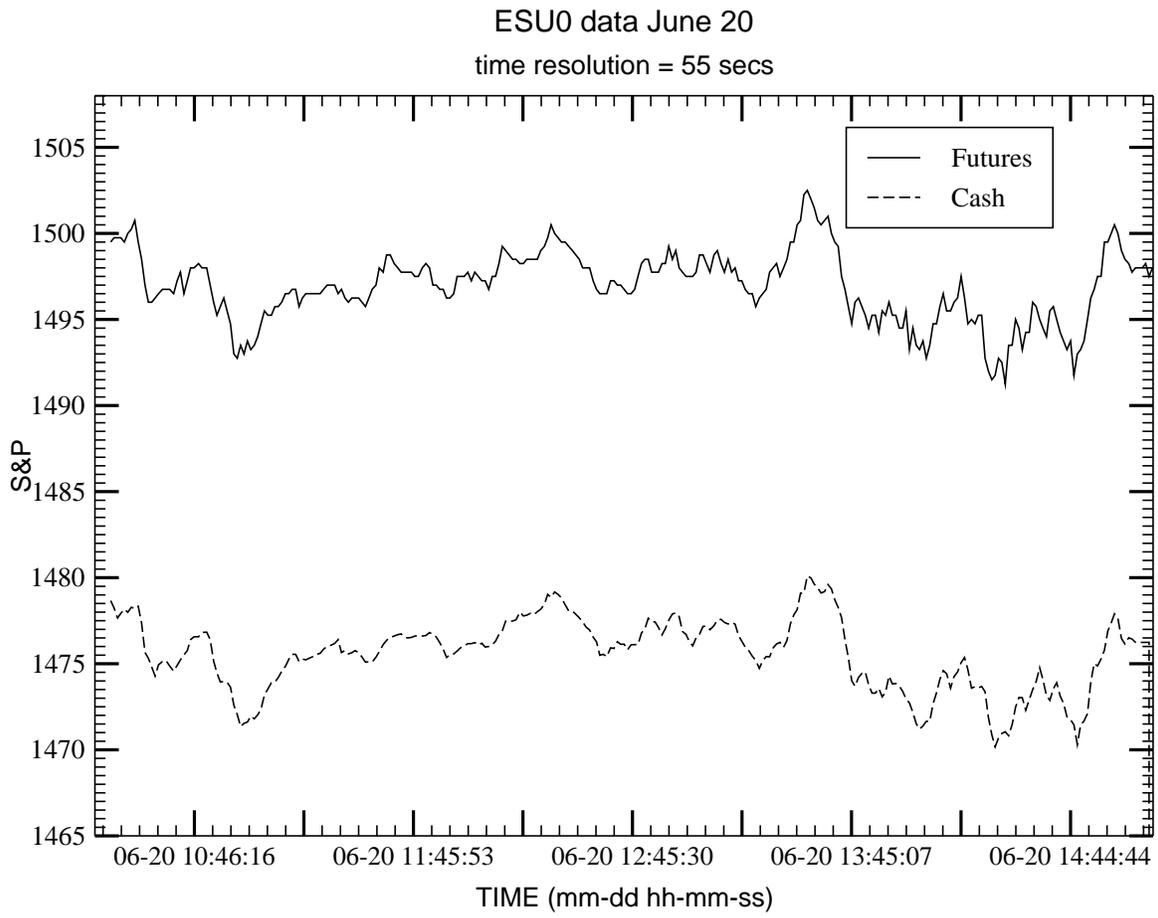

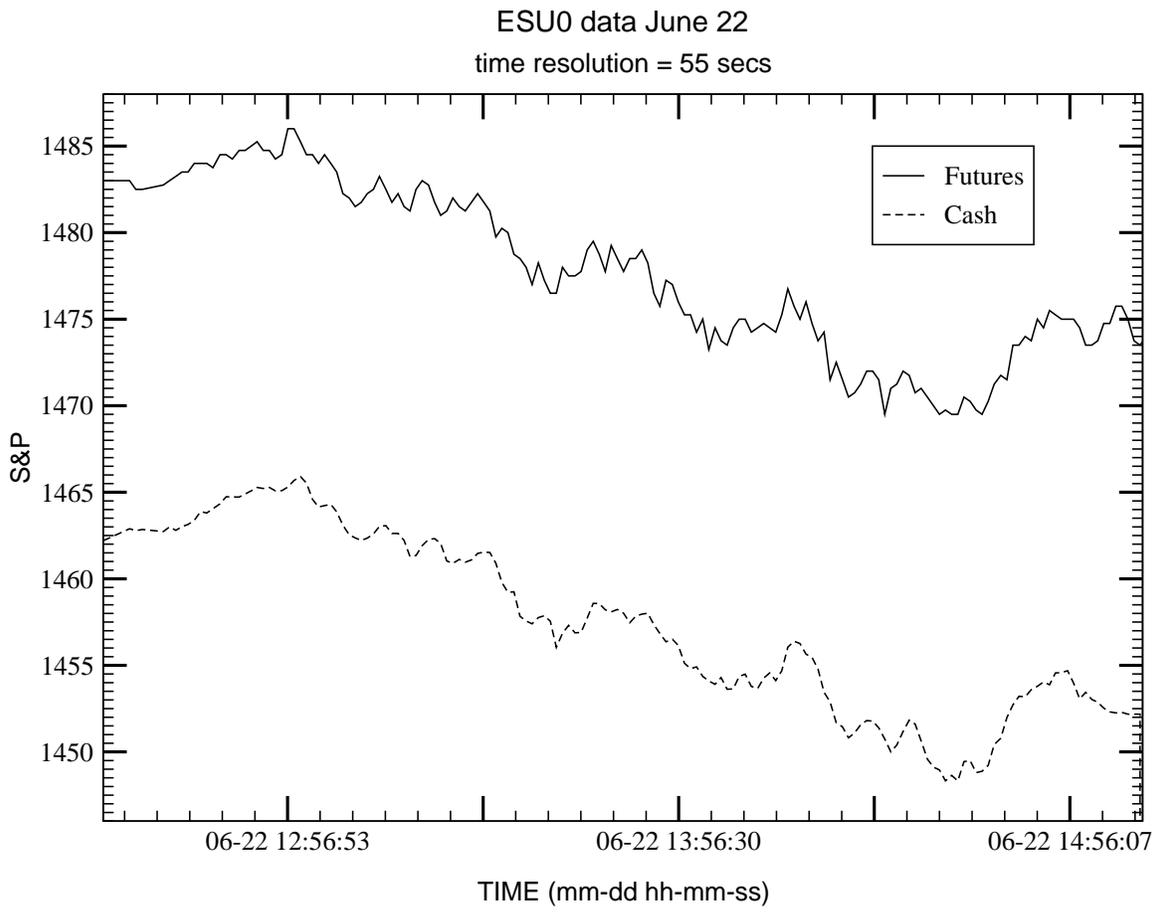

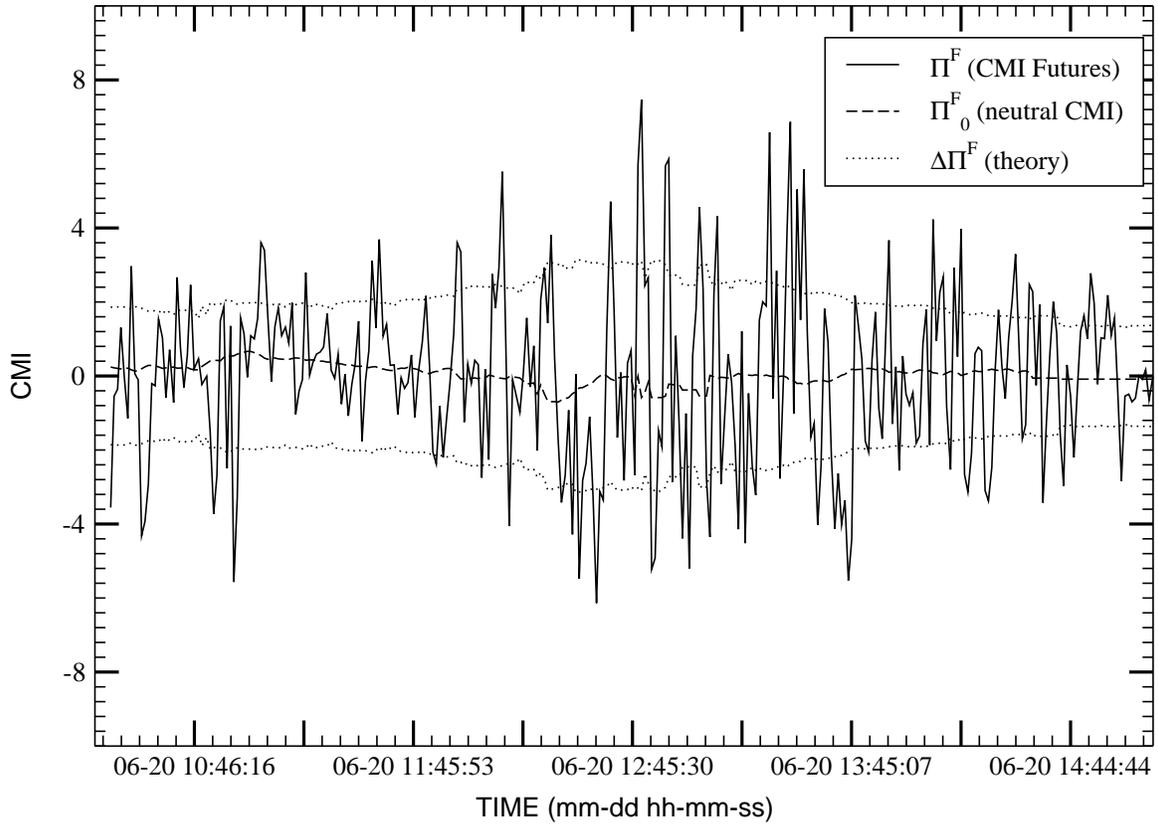

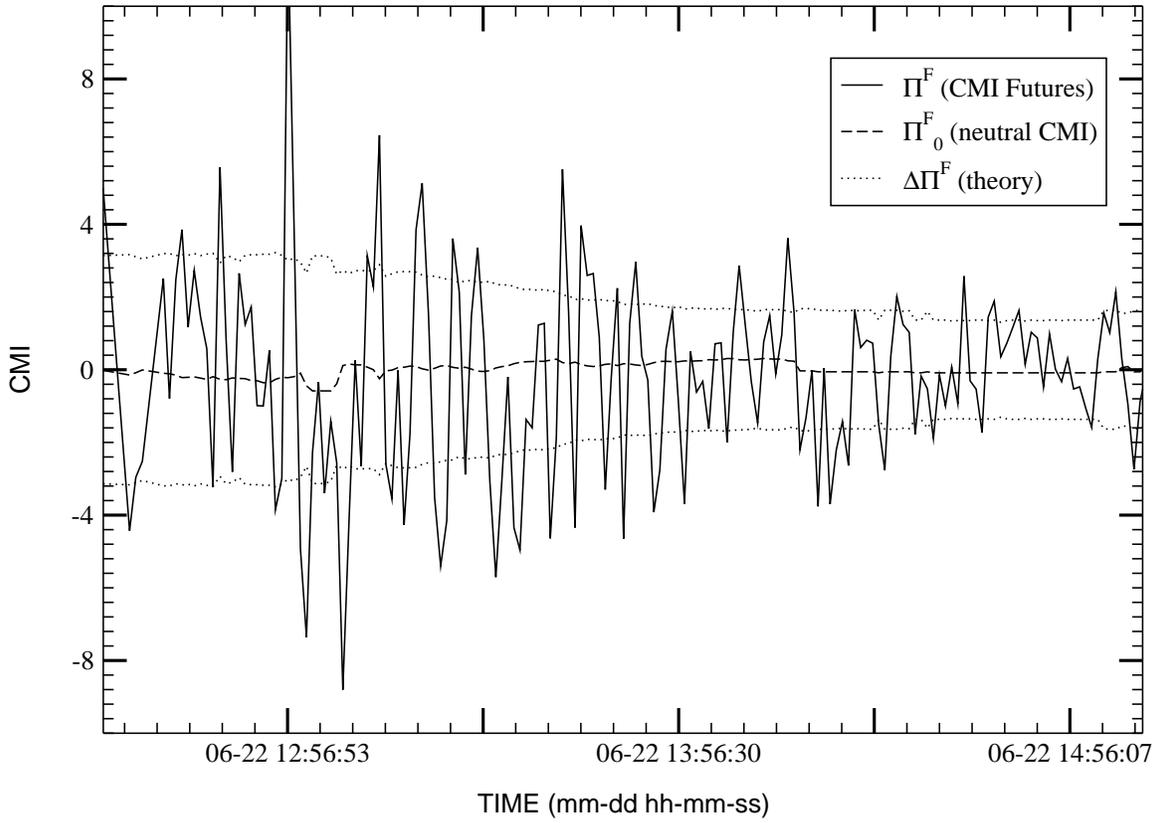

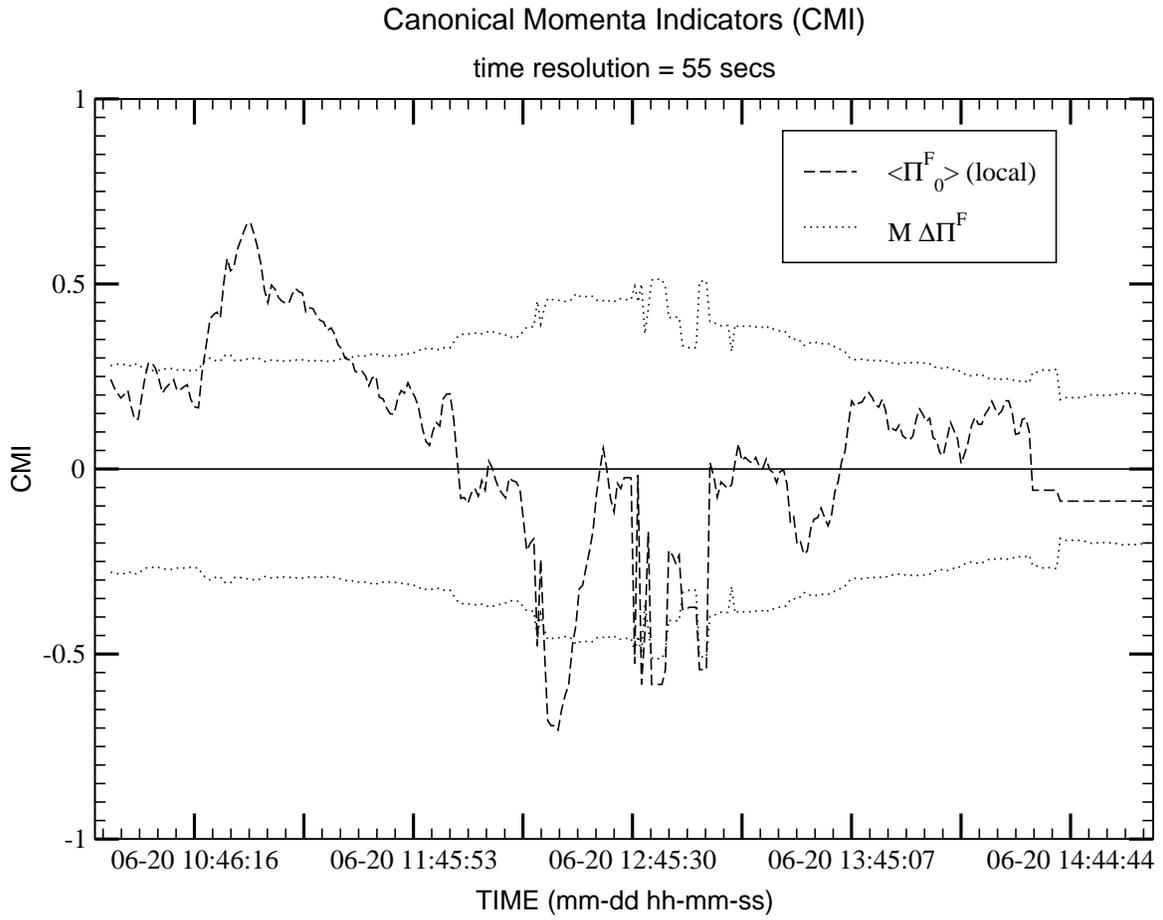

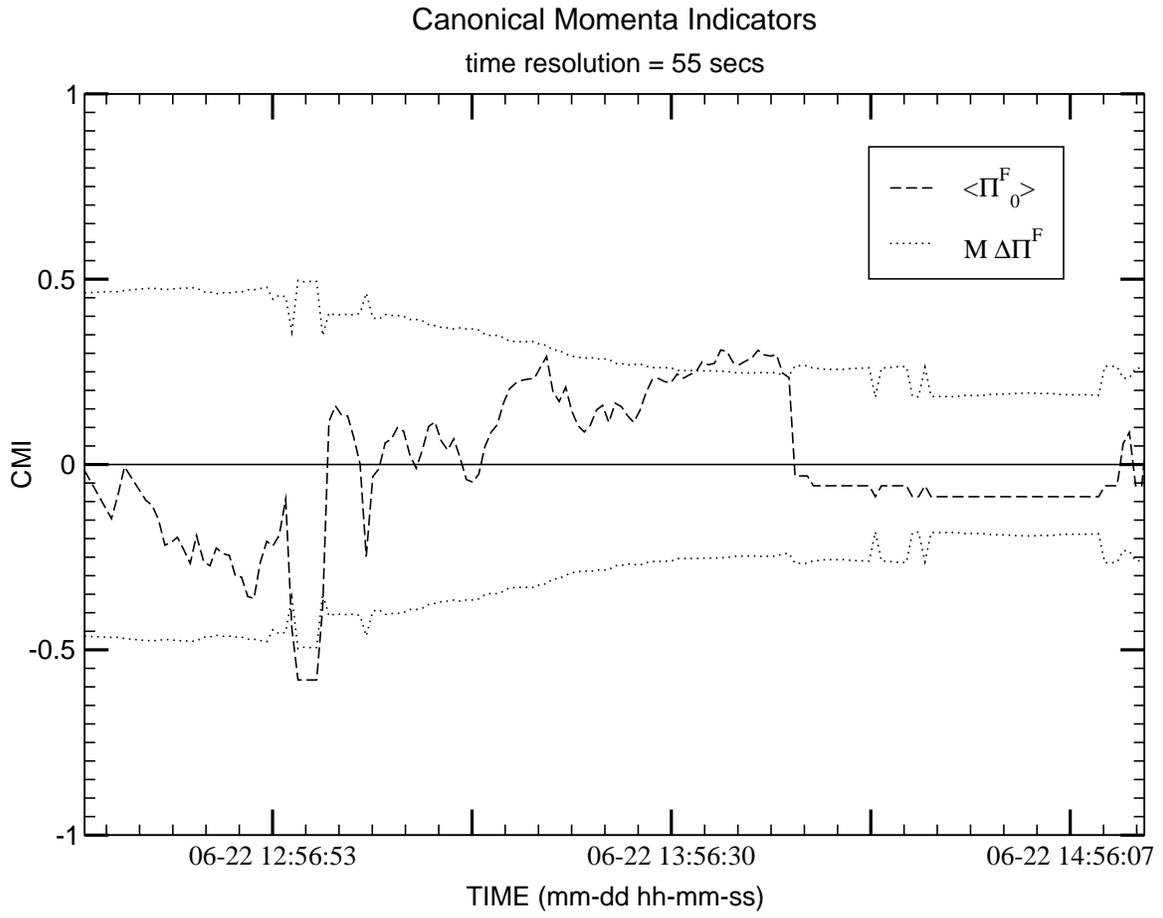

| TRAINING SET | TRADING RULES | STATISTICS | PARAMETERS ($\Delta t$ $W$ $M$) | | |
|---|---|---|---|---|---|
| 4D ESM0 0321-0324 | LOCAL MODEL | Parameters | 55 90 0.125 | 55 88 0.15 | 60 40 0.275 |
| | | $ profit (loss) | 1390 | 1215 | 1167 |
| | | # trades | 16 | 16 | 17 |
| | | # days + | 3 | 3 | 3 |
| | | % winners | 75 | 75 | 76 |
| | MULTIPLE MODELS | Parameters | 45 76 0.175 | 45 72 0.20 | 60 59 0.215 |
| | | $ profit (loss) | 2270 | 2167.5 | 1117.5 |
| | | # trades | 18 | 17 | 17 |
| | | # days + | 4 | 4 | 3 |
| | | % winners | 83 | 88 | 76 |
| 5D ESM0 0327-0331 | LOCAL MODEL | Parameters | 20 22 0.60 | 20 24 0.55 | 10 54 0.5 |
| | | $ profit (loss) | 437 | 352 | (35) |
| | | # trades | 15 | 16 | 1 |
| | | # days + | 3 | 3 | 0 |
| | | % winners | 67 | 63 | 0 |
| | MULTIPLE MODELS | Parameters | 45 74 0.25 | 40 84 0.175 | 30 110 0.15 |
| | | $ profit (loss) | 657.5 | 635 | 227.5 |
| | | # trades | 3 | 19 | 26 |
| | | # days + | 5 | 3 | 2 |
| | | % winners | 100 | 68 | 65 |
| 5D ESM0 0410-0414 | LOCAL MODEL | Parameters | 50 102 0.10 | 50 142 0.10 | 35 142 0.10 |
| | | $ profit (loss) | 1875 | 1847 | 1485 |
| | | # trades | 35 | 19 | 34 |
| | | # days + | 3 | 3 | 4 |
| | | % winners | 60 | 58 | 62 |
| | MULTIPLE MODELS | Parameters | 45 46 0.25 | 40 48 0.30 | 60 34 0.30 |
| | | $ profit (loss) | 2285 | 2145 | 1922.5 |
| | | # trades | 39 | 23 | 29 |
| | | # days + | 3 | 3 | 3 |
| | | % winners | 72 | 87 | 72 |

| TESTING SETS | STATISTICS | PARAMETERS ($\Delta t\ W\ M$) | | | | | |
|---|---|---|---|---|---|---|---|
| | | LOCAL MODEL | | | MULTIPLE MODELS | | |
| | | 55 90 0.125 | 55 88 0.15 | 60 40 0.275 | 45 76 0.175 | 45 72 0.20 | 60 59 0.215 |
| 5D ESM0 0327-0331 | $ profit (loss) | (712) | (857) | (1472) | (605) | (220) | (185) |
| | # trades | 20 | 17 | 16 | 18 | 12 | 11 |
| | # days + | 2 | 2 | 1 | 3 | 1 | 1 |
| | % winners | 50 | 47 | 44 | 67 | 67 | 54 |
| 4D ESM0 0403-0407 | $ profit (loss) | (30) | 258 | 602 | 1340 | 2130 | 932 |
| | # trades | 18 | 13 | 16 | 16 | 17 | 13 |
| | # days + | 3 | 3 | 2 | 1 | 1 | 1 |
| | % winners | 56 | 54 | 56 | 50 | 53 | 38 |
| 5D ESM0 0410-0414 | $ profit (loss) | 750 | 1227 | (117) | (530) | (1125) | (380) |
| | # trades | 30 | 21 | 23 | 23 | 20 | 18 |
| | # days + | 3 | 3 | 3 | 2 | 2 | 3 |
| | % winners | 60 | 62 | 48 | 48 | 50 | 50 |